\documentclass[conference,9pt]{IEEEtran}
\IEEEoverridecommandlockouts

\usepackage{amsthm}
\usepackage{amsfonts}
\usepackage{amssymb,bm}
\usepackage{mathrsfs}
\usepackage{mathtools}
\usepackage{amsmath}

\usepackage{cite}
\usepackage{balance}

\usepackage{url}
\usepackage{graphicx}
\usepackage{textcomp}

\usepackage{accents}
    
\usepackage{xcolor}

\def\BibTeX{{\rm B\kern-.05em{\sc i\kern-.025em b}\kern-.08em
    T\kern-.1667em\lower.7ex\hbox{E}\kern-.125emX}}
    
\begin{document}
\title{Multi-target Range and Angle detection for MIMO-FMCW radar with limited antennas}
\author{\IEEEauthorblockN{Himali Singh and
Arpan Chattopadhyay}
\IEEEauthorblockA{Electrical Engineering Department, Indian Institute of Technology Delhi, India. Email: \{eez208426, arpanc\}@ee.iitd.ac.in.}
\thanks{A. C. acknowledges support via the faculty seed grant, professional development fund and professional development  allowance from IIT Delhi, and grant no. RP04215G from I-Hub Foundation for Cobotics. H. S. acknowledges support via Prime Minister Research Fellowship.}
}
\maketitle

\begin{abstract}
Multiple-input multiple-output (MIMO) radar has several advantages with respect to the traditional radar array systems in terms of performance and flexibility. However, in order to achieve high angular resolution, a MIMO radar requires a large number of transmit/ receive antennas, which increases hardware design and computational complexities. Although spatial compressive sensing (CS) has been recently considered for a pulsed-waveform MIMO radar with sparse random arrays, such methods for the frequency-modulated continuous wave (FMCW) radar remain largely unexplored. In this context, we propose a novel multi-target localization algorithm in the range-angle domain for a MIMO FMCW radar with a sparse array of randomly placed transmit/ receive elements. In particular, we first obtain the targets’ range-delays using a discrete Fourier transform (DFT)-based focusing operation. The target angles are then recovered at each detected range using CS-based techniques exploiting the sparsity of the target scene. Our simulation results demonstrate the effectiveness of the proposed algorithm over the classical methods in detecting multiple targets with a sparse array.
\end{abstract}


\section{Introduction}
\label{sec:intro}
Frequency-modulated continuous wave (FMCW) radars have become a popular choice for short-range applications like automotive radars\cite{sun2020mimo,patole2017automotive}, human vital sign monitoring\cite{xu2022simultaneous}, synthetic aperture radars (SARs)\cite{meta2007signal}, and surveillance systems\cite{saponara2017radar}. The main advantages of FMCW radar are portability, low cost, and high resolution. An FMCW radar transmits a finite train of (piece-wise) linear frequency-modulated (LFM) chirps in each coherent processing interval (CPI). At the receiver, the target returns are mixed with the transmitted signal to obtain a complex sinusoidal beat or intermediate frequency (IF) signal. The targets' locations (and velocities if moving) information can be extracted from the frequencies of this IF signal. To this end, fast Fourier transforms (FFTs) have   traditionally been used to estimate the IF signal frequencies\cite{sun2020mimo}. However, to localize targets in the angular domain, multiple transmit and receive antennas are required. In MIMO radars, multiple orthogonal waveforms are transmitted simultaneously with the target returns processed jointly by the multiple receive antennas. The MIMO radar achieves a better angular resolution than conventional radar by exploiting a large number of degrees of freedom of a virtual array synthesized with a small number of physical antenna elements. In this work, we focus on multi-target range-angle detection using MIMO FMCW radars. Conventionally, two-dimensional frequency estimation algorithms are used to estimate both targets' ranges and angles of arrival (AOAs) from the received signal. Other frequency estimation algorithms considered for MIMO FMCW radar include 2D-FFT\cite{feger200977}, 2D-MUSIC\cite{belfiori20122d}, and ESPRIT\cite{lemma2003analysis}.

From the array processing theory, it is known that a high angular resolution requires a large array aperture\cite{richards2014fundamentals}. Further, increasing the aperture without a parallel increase in antenna elements leads to ambiguities in angle estimation. Although MIMO technology helps to achieve higher resolution, the cost of synthesizing a large virtual array with the half-wavelength element spacing (spatial Nyquist sampling rate) can be very high. In this context, sparse linear arrays (SLAs) have been proposed recently for both pulsed-waveform and continuous-wave radars\cite{rossi2013spatial,feger200977,sun2020sparse}. Optimal sparse array design was considered in \cite{diamantaras2021sparse} while \cite{feger200977} designed a non-uniform SLA and applied digital beamforming techniques for AOA estimation after interpolating for the missing measurements in the synthesized SLA. On the other hand, \cite{sun2020sparse} suggested matrix completion techniques to complete the corresponding linear array for angle detection.

Compressed sensing (CS) addresses sparse signal recovery with fewer measurements\cite{elad2010sparse}. The sparse array setup enables spatial compressive sensing such that the CS recovery naturally suits our target localization problem. Note that the target scene is sparse since only a small number of targets are present in the scene. The CS-recovery-based localization has recently been applied for angle estimation for pulsed-MIMO radar\cite{rossi2013spatial}. In \cite{alistarh2022compressed}, CS-based algorithms were used to process measurements from a traditional full array. Besides, spatial compression, CS techniques have also been considered in radars for reduced sampling rate\cite{bar2014sub,yu2010mimo}, interference mitigation\cite{correas2019sparse}, and multi-target shadowing effect mitigation in constant false-alarm rate (CFAR) detection\cite{cao2021compressed}.

\textbf{Contributions:} In this paper, we present a novel multi-target localization algorithm to detect targets' ranges and AOAs using a random SLA. Prior methods employing CS-based techniques (e.g. \cite{rossi2013spatial}) often address only angle detection at a prior known range bin. Here, we consider both range and angle detection in a MIMO FMCW radar. For range detection, we exploit a discrete Fourier transform (DFT)-based focusing operation followed by binary integration\cite{richards2014fundamentals} of measurements across pulses and virtual array channels, trading off range resolution for higher detection probability. For angle recovery, we use CS-based techniques, which relax the dependence of the angular resolution on the number of antenna elements. Finally, we illustrate the proposed method's performance through numerical simulations, comparing it with classical-FFT processing.

The rest of the paper is organized as follows. In the next section, we describe the FMCW radar system model with the random sparse MIMO array setup. In Section~\ref{sec:recovery algorithm}, we present the proposed range and angle detection algorithm. The simulation results are discussed in Section~\ref{sec:simulation}, followed by conclusins in Section~\ref{sec:summary}.

\section{Radar System model}
\label{sec:system model}
\begin{figure}
  \centering
  \includegraphics[width = 0.85\columnwidth]{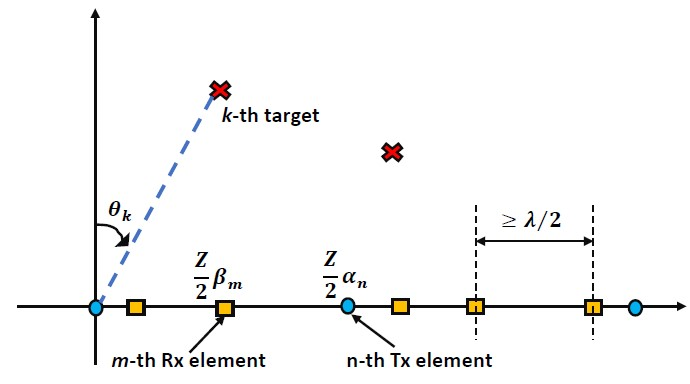}
  \caption{MIMO radar system ($\square$ and $\circ$ denote receivers and transmitters, respectively).}
 \label{fig:mimo radar}
\end{figure}

Consider a colocated MIMO radar system, as shown in Fig.~\ref{fig:mimo radar}, composed of $N_{T}$ transmitters and $N_{R}$ receivers forming a (possibly overlapping) array of total aperture $Z_{T}$ and $Z_{R}$, respectively, and define $Z\doteq Z_{T}+Z_{R}$. The $n$-th transmitter's and $m$-th receiver's locations along the x-axis are $Z\alpha_{n}/2$ and $Z\beta_{m}/2$, respectively, where $\alpha_{n}\in[ -Z_{T}/Z, Z_{T}/Z]$ and $\beta_{m}\in[ -Z_{R}/Z, Z_{R}/Z]$. Note that $\alpha_{n}$ and $\beta_{m}$ are randomly drawn from appropriate uniform distributions\cite{rossi2013spatial}. The transmitters transmit LFM chirps, orthogonal across transmitters. Consider $f_{c}$ as the carrier frequency and $\gamma$ as the chirp rate of the LFM chirp with chirp duration $T$. The FMCW radar's transmitted chirp is modeled as
\par\noindent\small
\begin{align*}
     s(\overline{t})=\exp{\left(j2\pi\left(f_{c}\overline{t}+\frac{\gamma}{2}\overline{t}^{2}\right)\right)},\;\;\; 0\leq \overline{t}\leq T,
\end{align*}
\normalsize
with $\overline{t}$ as the continuous-time index. A total of $P$ chirps is transmitted in each CPI. Different orthogonal waveform designs for MIMO-FMCW radar transmitters have been proposed in \cite{de2011orthogonal,babur2013nearly}. For simplicity, we consider time-domain multiplexing, where the transmitters transmit the same signal with relative time shifts. In our proposed detection algorithm, we process each transmitted chirp independently and use binary integration\cite{richards2014fundamentals} after detection across pulses (in a CPI) to obtain the estimated ranges. On the contrary, classical-FFT processing considers coherent or non-coherent integration of the pulses to average out the interference and noise before detection\cite{richards2014fundamentals}. In Section~\ref{sec:simulation}, we discuss how binary integration improves the detection probability over classical processing. Similarly, the orthogonality of the transmitted signals allows the corresponding received signal components to be separated at each receiver. Hence, we first focus on the received signal component at the $m$-th receiver due to the single chirp transmitted from the $n$-th transmitter.

We assume a target scene of $K$ stationary, far-field, non-fluctuating point targets (Swerling Case 0 model\cite{richards2014fundamentals}). We denote the $k$-th target's range and angle of arrival (AOA) as $R_{k}$ and $\theta_{k}$, respectively. Denote $\tau_{m,n,k}$ as the total time-delay in the $k$-th target's return at the $m$-th receiver from the $n$-th transmitted signal such that the received signal component is given as
\par\noindent\small
\begin{align*}
    r_{m,n}(\overline{t})=\sum_{k=1}^{K}a_{k}s(\overline{t}-\tau_{m,n,k}),
\end{align*}
\normalsize
where $a_{k}$ is the complex amplitude proportional to the $k$-th target's radar cross-section (RCS). The time delay $\tau_{m,n,k}$ consists of the range delay $\tau^{R}_{k}$ and angular delay $\tau^{\theta}_{m,n,k}$ as
\par\noindent\small
\begin{align}
    \tau_{m,n,k}=\tau^{R}_{k}+\tau^{\theta}_{m,n,k},\label{eqn:total delay}
\end{align}
\normalsize
where $\tau^{R}_{k}=2R_{k}/c$ and $\tau^{\theta}_{m,n,k}=Z(\alpha_{n}+\beta_{m})\sin{(\theta_{k})/2c}$ with constant $c$ denoting the speed of light. Note that the far-field assumption leads to a constant AOA across the array.

After mixing the $m$-th received signal with the $n$-th transmitted signal, the FMCW radar's IF signal $y_{m,n}(\overline{t})$ is represented as
\par\noindent\small
\begin{align*}
    y_{m,n}(\overline{t})&=\sum_{k=1}^{K}a_{k}^{*}\exp{\left(j2\pi\left(\gamma\tau_{m,n,k}\overline{t}+f_{c}\tau_{m,n,k}-\frac{\gamma}{2}\tau_{m,n,k}^{2}\right)\right)}\\
    &\;\;\;+w_{m,n}(\overline{t}),
\end{align*}
\normalsize
where $(\cdot)^{*}$ represents the conjugate operation and $w_{m,n}(\overline{t})$ is the interference plus-noise term. Each IF signal $y_{m,n}(\overline{t})$ is sampled at sampling frequency $f_{s}$ as
\par\noindent\small
\begin{align*}
    y_{m,n}[t]&=\sum_{k=1}^{K}a_{k}^{*}\exp{\left(j2\pi\left(\gamma\tau_{m,n,k}\frac{t}{f_{s}}+f_{c}\tau_{m,n,k}-\frac{\gamma}{2}\tau_{m,n,k}^{2}\right)\right)}\\
    &\;\;\;+w_{m,n}[t],
\end{align*}
\normalsize
for $0\leq t\leq N-1$, where $N=f_{s}T$ is the total number of samples in a single pulse and $w_{m,n}[t]$ is the sampled noise. Here, we represent the discrete-time index by $t$. For the $N_{T}$ transmitters and $N_{R}$ receivers MIMO setup, we obtain `$N_{T}N_{R}$' sampled measurements $\{y_{m,n}[t]\}_{1\leq m\leq N_{R},1\leq n\leq N_{T}}$ for all $P$ pulses.

\section{Sparse array Recovery algorithm}
\label{sec:recovery algorithm}
In this section, we describe the proposed range-angle detection algorithm. The spatial compressive sensing framework proposed in \cite{rossi2013spatial} for pulsed MIMO radar assumes an independent range-Doppler processing and focus only on targets in a given range-Doppler bin for AOA estimation. On the contrary, here, we consider both range and AOA detection. In Section~\ref{subsec:range detection}, we adopt a DFT-focusing operation to estimate the targets' ranges and separate the range and AOA information. Finally, in Section~\ref{subsec:angle detection}, the CS-based recovery provides the AOA estimates at each detected range bin.

\subsection{Range detection}
\label{subsec:range detection}
Consider the $N$-point DFT of the sampled IF signal $y_{m,n}[t]$ as
\par\noindent\small
\begin{align}
    &Y_{m,n}[l]=\sum_{t=0}^{N-1}y_{m,n}[t]\exp{(-j2\pi lt/N)},\nonumber\\
    &=\sum_{k=1}^{K}a^{*}_{k}\exp{\left(j2\pi\left(f_{c}\tau_{m,n,k}-\frac{\gamma}{2}\tau_{m,n,k}^{2}\right)\right)}\nonumber\\
    &\;\;\;\times\sum_{t=0}^{N-1}\exp{\left(j2\pi\left(\frac{\gamma\tau_{m,n,k}}{f_{s}}-\frac{l}{N}\right)t\right)}+W_{m,n}[l],\label{eqn:DFT of y}
\end{align}
\normalsize
for $0\leq l\leq N-1$, where $W_{m,n}[l]=\sum_{t=0}^{N-1}w_{m,n}[t]\exp{(-j2\pi lt/N)}$ represents the noise term.

Replacing $N=f_{s}T$, we first analyze the sum of exponents $\sum_{t=0}^{N-1}\exp{\left(j(\frac{2\pi\gamma}{f_{s}})\left(\tau_{m,n,k}-\frac{l}{\gamma T}\right)t\right)}$ in \eqref{eqn:DFT of y}. Consider the sum of $M$ exponents $g(x|\overline{x})=\sum_{q=0}^{M-1}e^{j(x-\overline{x})q\omega}$ for given constants $\overline{x}$ and $\omega$. We can approximate $|g(x|\overline{x})|$ as
\par\noindent\small
\begin{align*}
    |g(x|\overline{x})|=\begin{cases}M,&|x-\overline{x}|\leq\pi/M\omega\\0,&|x-\overline{x}|>\pi/M\omega\end{cases}.
\end{align*}
\normalsize
The approximation implies that in the focus zone $|x-\overline{x}|\leq\pi/M\omega$, the $M$ exponents are coherently integrated while the signal outside the focus zone is severely attenuated. In \cite{bar2014sub}, this focusing approximation was introduced as Doppler focusing across pulses in a CPI to reduce the joint delay-Doppler estimation problem to delay only estimation at a particular Doppler frequency. In our case, the sum of exponents appears naturally in the DFT of $y_{m,n}[t]$.

Using the focusing approximation for the sum of $N$ exponents (indexed by $t$) in \eqref{eqn:DFT of y}, we have
\par\noindent\small
\begin{align}
    Y_{m,n}[l]\approx\sum_{k'=1}^{K'}a^{*}_{k'}N\exp{\left(j2\pi\left(f_{c}\tau_{m,n,k'}-\frac{\gamma}{2}\tau_{m,n,k'}^{2}\right)\right)}+W_{m,n}[l],\label{eqn:Ymn}
\end{align}
\normalsize
where $\{a_{k'},\tau_{m,n,k'}\}_{1\leq k'\leq K'}$ represents the subset of targets which satisfy $|\tau_{m,n,k'}-l/(\gamma T)|\leq 1/(2\gamma T)$ for the given $l$-th DFT bin. Assuming $\tau^{R}_{k}\gg\tau^{\theta}_{m,n,k}$ for all targets, we have $\tau_{m,n,k'}\approx\tau^{R}_{k'}$ such that the received signal from targets at ranges satisfying $|\tau^{R}_{k'}-l/(\gamma T)|\leq 1/(2\gamma T)$ are coherently integrated, resulting in a (magnitude) peak at the $l$-th DFT bin. Furthermore, the practical values of $\gamma$ and $T$ for an FMCW radar ensures that the value $1/(2\gamma T)$ is small enough and $\tau^{R}_{k'}\approx l/(\gamma T)$. Hence, using threshold detection to identify the peaks in $Y_{m,n}[l]$ (corrupted by noise), we obtain the range estimates. The estimated range $R'$ corresponding to a DFT peak at $l'$-th bin is computed as
\par\noindent\small
\begin{align*}
    R'=\frac{c l'}{2\gamma T}.
\end{align*}
\normalsize
These range estimates are computed independently for all $P$ pulses and for all $N_{T}N_{R}$ measurements $\{y_{m,n}[t]\}_{1\leq m\leq N_{R},1\leq n\leq N_{T}}$. The detected ranges are first filtered for false alarms across the $P$ pulses using binary integration, i.e., only the ranges detected in a sufficient number of pulses are considered valid target ranges. Similarly, the detected ranges are also filtered across the $N_{T}N_{R}$ measurements which further reduces the false alarm probability.

The classical-FFT range processing also involves threshold detection for peaks in the DFT of the sampled IF signal. However, in classical processing, all the pulses are processed together non-coherently to compute the DFT, which increases the range resolution by increasing the frequency resolution of the computed DFT. On the other hand, by processing each pulse independently, we trade off range resolution for reduced missed detection probability. In particular, in the case of close-range targets, the classical processing often suffers from false peaks dominating the actual target peaks. Using binary integration across pulses and then across $N_{T}N_{R}$ virtual array channels, the detection probability is enhanced with a constant false alarm probability. This performance improvement with binary integration is further discussed in Section~\ref{subsec:DFT processing} with a simulated example of three close-range targets.

\subsection{Angle detection}
\label{subsec:angle detection}
Consider a detected range bin at the $l'$-th DFT point. Substituting \eqref{eqn:total delay} in \eqref{eqn:Ymn} for $\tau_{m,n,k'}$, we obtain
\par\noindent\small
\begin{align*}
     &Y_{m,n}[l']=W_{m,n}[l']\\
     &+\sum_{k'=1}^{K'}a^{*}_{k'}N\exp{\left(j2\pi\left(f_{c}\tau^{R}_{k'}-\frac{\gamma}{2}(\tau^{R}_{k'})^{2}\right)\right)}\exp{(j2\pi(f_{c}-\gamma\tau^{R}_{k'})\tau^{\theta}_{m,n,k'})},
\end{align*}
\normalsize
using $(\tau^{R}_{k'})^{2}\gg(\tau^{\theta}_{m,n,k'})^{2}$. For practical FMCW radars, carrier frequency $f_{c}$ (in GHz), chirp rate $\gamma$ (in MHz/$\mu$s) and short-range delay $\tau^{R}_{k}$ (a few $\mu$s) are such that the term $\gamma\tau^{R}_{k'}$ is negligible and
\par\noindent\small
\begin{align}
     Y_{m,n}[l']&=\sum_{k'=1}^{K'}a^{*}_{k'}N\exp{\left(j2\pi\left(f_{c}\tau^{R}_{k'}-\frac{\gamma}{2}(\tau^{R}_{k'})^{2}\right)\right)}\exp{(j2\pi f_{c}\tau^{\theta}_{m,n,k'})}\nonumber\\
     &\;\;\;+W_{m,n}[l'].\label{eqn:Ymn for angle}
\end{align}
\normalsize
Note that the exponential terms with the range and angle delays are now separated in $Y_{m,n}[l']$.

Denote $x_{k}\doteq a^{*}_{k}N\exp{\left(j2\pi\left(f_{c}\tau^{R}_{k}-\frac{\gamma}{2}(\tau^{R}_{k})^{2}\right)\right)}$ as the complex amplitude independent of the AOAs. Further, we denote $Y^{p}_{m,n}[l']$ as the $l'$-th DFT coefficient computed for the $p$-th pulse. Stack the measurements $Y^{p}_{m,n}[l']$ for all $(m,n)$-pairs in a $N_{T}N_{R}\times 1$ vector $\mathbf{y}_{p}$. Now, define the `$N_{T}N_{R}\times P$' matrix $\mathbf{Y}=[\mathbf{y}_{1},\hdots,\mathbf{y}_{p}]$. Similarly, define the $K'\times P$ matrix $\widetilde{\mathbf{X}}=[\widetilde{\mathbf{x}}_{1},\hdots,\widetilde{\mathbf{x}}_{P}]$ with $\widetilde{\mathbf{x}}_{p}=[x_{1},\hdots,x_{K'}]^{T}$. Now, substituting $\tau^{\theta}_{m,n,k}=Z(\alpha_{n}+\beta_{m})\sin{(\theta_{k})/2c}$ in \eqref{eqn:Ymn for angle} yields
\par\noindent\small
\begin{align}
     \mathbf{Y}=\widetilde{\mathbf{C}}(\bm{\theta})\widetilde{\mathbf{X}}+\mathbf{W},\label{eqn:AOA matrix eqn}
\end{align}
\normalsize
where the $N_{T}N_{R}\times K'$ matrix $\widetilde{\mathbf{C}}(\bm{\theta})=[\mathbf{c}(\theta_{1}),\hdots,\mathbf{c}(\theta_{K'})]$ with each column 
\small
$\mathbf{c}(\theta)=[\exp{(j\pi f_{c}Z(\alpha_{1}+\beta_{1})sin(\theta))},\hdots,\exp{(j\pi f_{c}Z(\alpha_{N_{T}}+\beta_{N_{R}})sin(\theta))}]^{T}$,
\normalsize
known as the virtual array steering vector\cite{rossi2013spatial} parameterized by the AOA $\theta$. Here, $\mathbf{W}$ represents the $N_{T}N_{R}\times P$ noise matrix obtained from similarly stacking $W_{m,n}[l']$ from all pulses.

We need to recover $\bm{\theta}$ and $\widetilde{\mathbf{X}}$ from $\mathbf{Y}$ with a small number of antenna elements. To this end, we use a sparse localization framework. Assume a grid of $G$ points $\bm{\phi}_{1\leq g\leq G}$ of the possible target AOAs $\theta$ with $G\gg K$ and negligible discretization errors. Each grid element $\phi_{g}$ parameterizes a column of $\widetilde{\mathbf{C}}(\bm{\theta})$. Hence, we can define a $N_{T}N_{R}\times G$ dictionary matrix $\mathbf{C}=[\mathbf{c}(\phi_{1}),\hdots,\mathbf{c}(\phi_{G})]$. From \eqref{eqn:AOA matrix eqn}, the measurements $\mathbf{Y}$ are then expressed as
\par\noindent\small
\begin{align}
     \mathbf{Y}=\mathbf{C}\mathbf{X}+\mathbf{W},\label{eqn:AOA cs eqn}
\end{align}
\normalsize
where the unknown $G\times P$ matrix $\mathbf{X}$ contains the target AOAs and complex amplitudes ($x_{k}$). A non-zero row of $\mathbf{X}$ represents a target present at the corresponding grid point. Hence, the system \eqref{eqn:AOA cs eqn} is sparse since $\mathbf{X}$ has only $K'\ll G$ non-zero rows for a particular detected range bin. Given the measurements $\mathbf{Y}$ and matrix $\mathbf{C}$, AOA estimation reduces to determining the support (non-zero rows) of $\mathbf{X}$. Note that the matrix $\mathbf{C}$ and hence, the recovery guarantees depend on the choice of grid points $\bm{\phi}_{1\leq g\leq G}$ as well as the number and (random) positions of the transmitters and receivers ($\{\alpha_{n}\}_{1\leq n\leq N_{T}}$ and $\{\beta_{m}\}_{1\leq m\leq N_{R}}$). In \cite{rossi2013spatial}, authors also discuss the sufficient conditions on the grid and the random array for recovery of $\mathbf{X}$ with high probability.

For the recovery of sparse matrix $\mathbf{X}$ with limited antenna elements, we consider CS-based algorithms. CS problems can be classified as single measurement vector (SMV) models for $P=1$ where $\mathbf{Y}$ reduces to a single vector, or multiple measurement vector (MMV) models for $P\geq 1$. Our problem \eqref{eqn:AOA cs eqn} is an MMV setting. However, we first consider the SMV setting with $P=1$ such that $\mathbf{Y}=\mathbf{y}$, $\mathbf{X}=\mathbf{x}$ and $\mathbf{W}=\mathbf{w}$ in \eqref{eqn:AOA cs eqn}.

Recovering a sparse $\mathbf{x}$ from $N_{T}N_{R}$ measurements $\mathbf{y}$ involves solving the non-convex combinatorial $l_{0}$-norm problem
\par\noindent\small
\begin{align}
     \textrm{min}_{\mathbf{x}} \|\mathbf{x}\|_{0}\;\;\; \textrm{s.t.}\;\;\;\|\mathbf{y}-\mathbf{C}\mathbf{x}\|_{2}\leq\nu,\label{eqn:CS problem}
\end{align}
\normalsize
where parameter $\nu$ is chosen based on the noise level $\|\mathbf{w}\|_{2}$ or the sparsity of $\mathbf{x}$. Solution of \eqref{eqn:CS problem} requires an exhaustive search of exponential complexity\cite{elad2010sparse}. However, an approximate solution can be obtained using a variety of polynomial complexity algorithms. Matching pursuit (MP) is one such family of methods, which iteratively refines the provisional support by adding one dictionary element at a time. Orthogonal MP (OMP)\cite{pati1993orthogonal}, orthogonal least squares (OLS)\cite{chen1989orthogonal}, and compressive sampling MP (CoSaMP)\cite{needell2009cosamp} are some popular MP algorithms for the SMV setting. For the general MMV setting, simultaneous OMP (SOMP)\cite{tropp2006algorithms} extends the OMP algorithm to matrix measurements. Another class of recovery algorithms is the Basis pursuit (BP) which relaxes the $l_{0}$-norm in \eqref{eqn:CS problem} with $l_{1}$-norm, resulting in a convex problem whose global solution can be found in polynomial time\cite{candes2008introduction}. In Section~\ref{subsec:performance}, we consider OMP and SOMP, respectively, for the sparse recovery in SMV and MMV settings.

\section{Simulation results}
\label{sec:simulation}
We now demonstrate the performance of the proposed method in comparison to the classical FFT-processing. In Section~\ref{subsec:DFT processing}, we first investigate the effect of binary integration for range processing discussed in Section~\ref{subsec:range detection}. The simulation results for a sparse target scene are provided in Section~\ref{subsec:performance}.

We considered a MIMO-FMCW radar system transmitting at carrier frequency $f_{c}=9.4$ GHz. The transmitted bandwidth was chosen as $B=250$ MHz with chirp duration $T=363 \mu$s (chirp rate $\gamma=B/T$) and sampling frequency $f_{s}=1.4$ MHz such that the range resolution was $0.6$ m. One CPI consisted of $P=10$ MIMO sweeps. For the sparse array, $3$ transmitters and $3$ receivers (total $6$ antenna elements) were placed uniformly over the array apertures $Z_{T}=Z_{R}=6\lambda$, where $\lambda$ is the wavelength of the transmitted signal. Note that in this case $\alpha_{n},\beta_{m}\in[-0.5,0.5]$ for $1\leq n,m\leq 3$. For the full array, we considered $4$ transmitters and $8$ receivers arranged as in \cite{belfiori20122d}. In particular, two transmitters were placed on either side of the array with an inter-element spacing of $\lambda$. The receivers were placed in the middle with an inter-element spacing of $0.5\lambda$ and $0.25\lambda$ spacing between the closest transmitter-receiver elements. This arrangement results in a virtual array of $20$ unique element locations with $0.5\lambda$ uniform separation. The target gains were generated as $a_{k}=\exp{(j\psi_{k})}$ with $\psi_{k}$ drawn from i.i.d. uniform distribution over $[0,2\pi)$. The noise term $w_{m,n}[t]$ is modeled as i.i.d. zero-mean complex circular Gaussian noise $\mathcal{CN}(0,\sigma^{2}\mathbf{I})$, mutually independent across pulses and virtual array channels. The signal-to-noise ratio (SNR) is then defined as $-10\log_{10}{(\sigma^{2})}$\cite{rossi2013spatial}.

\subsection{DFT processing: classical and proposed method}
\label{subsec:DFT processing}
\begin{figure}
  \centering
  \includegraphics[width = \columnwidth]{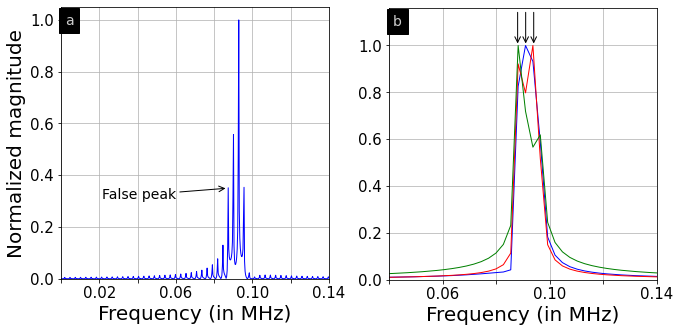}
  \caption{Normalized DFT magnitude for (a) Classical range-FFT; and (b) Three different pulses for the proposed method (arrows indicate the detected peaks).}
 \label{fig:DFT}
\end{figure}
Consider three close-range targets with ranges $R_{1}=20.6$ m, $R_{2}=20.0$ m and $R_{3}=19.4$ m at AOAs $\theta_{1}=\theta_{2}=\theta_{3}=0^{\circ}$. Considering the noise-free case, Fig.~\ref{fig:DFT}a shows the range-FFT computed in the classical-FFT processing. Fig.~\ref{fig:DFT}b shows the DFT computed in the proposed method for three different pulses from measurement $y_{1,1}[t]$, which are then used to estimate the target ranges using binary integration as detailed in Section~\ref{subsec:range detection}. We observe that non-coherent processing of the pulses in the classical method provides a refined spectrum as compared to the proposed method of processing one pulse at a time. However, the classical range-FFT suffers from side-lobe effect which results in a false peak of the same order of the third target ($R_{3}$) peak. Hence, reducing the false alarms (increasing the threshold) results in a missed detection. On the other hand, in the proposed binary integration method, the missed targets in one pulse can be detected at other pulses (or some other $y_{m,n}[t]$ measurement). Hence, binary integration can enhance the detection probability for a constant false alarm rate by trading off range resolution.

\subsection{Performance analysis}
\label{subsec:performance}
\begin{figure}
  \centering
  \includegraphics[width = \columnwidth]{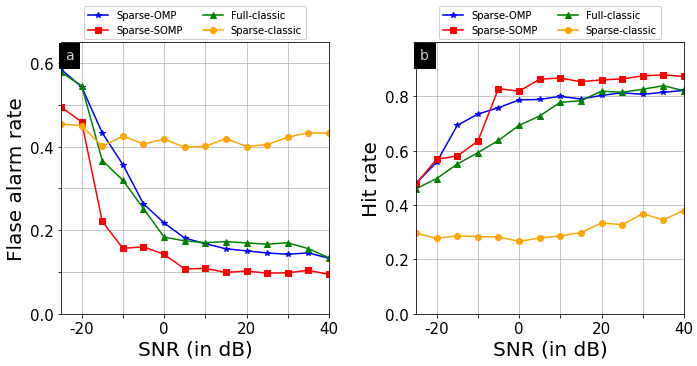}
  \caption{Average (a) false alarm rate, and (b) hit rate at different SNRs for classical-FFT processing and the proposed method.}
 \label{fig:rates}
\end{figure}
\begin{figure}
  \centering
  \includegraphics[width = \columnwidth]{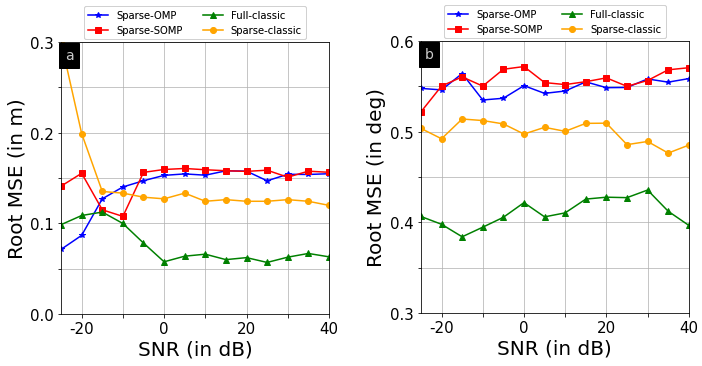}
  \caption{Root MSE in (a) range, and (b) angle estimation at different SNRs for classical-FFT processing and the proposed method.}
 \label{fig:error}
\end{figure}
We considered $K=5$ targets with target delays and AOAs chosen uniformly at random with ranges in $[10 m, 40 m]$ and AOAs in $[-15^{\circ},15^{\circ}]$. This ensures the target scene has close-range targets as well as multiple targets at the same range. For the proposed CS-based angle recovery, we considered OMP with the vector measurement $\mathbf{y}$ as the sum across $10$ pulses, and SOMP for matrix measurement $\mathbf{Y}$. In \cite{rossi2013spatial}, authors assumed a known sparsity level and used the prior information of the actual number of targets $K$ in the CS algorithms. On the contrary, here, we assumed a sparsity level of $K_{max}$ for OMP and SOMP algorithms. The target AOAs were then obtained using threshold detection on the recovered signal. Hence, we do not require a prior estimate of $K$. We set $K_{max}=10$ for both OMP and SOMP. The grid $\bm{\phi}_{1\leq g\leq G}$ was chosen as $150$ uniformly spaced points in the $\sin(\theta)$ domain in the interval $[-0.7071,0.7071]$. Note that the AOA estimates are uniform in the $\sin{(\theta)}$ domain. This holds for classical-FFT processing as well where the DFT samples are equally spaced in the $\sin{(\theta)}$ domain between $[-1,1]$ and assume a non-linear distribution in the $\theta$ domain. The grid $\bm{\phi}_{1\leq g\leq G}$ spans the interval $[-45^{\circ},45^{\circ}]$ in the AOA domain.

We consider hit rate and root-mean-squared error (RMSE) of the recovered targets as the performance metrics. A `hit' is defined as a range-angle estimate within $0.6$ m in range and $1^{\circ}$ in angle of the true target. The recovery error is computed for the estimates classified as hits. The target estimates not classified as hits are the false alarms. We vary the thresholds of the threshold detectors to maintain a constant false-alarm rate at different SNRs. The hit rate and false alarm rate for different SNRs, averaged over $300$ independent simulations, are shown in Fig.~\ref{fig:rates} for the proposed method and classical-FFT processing considering both full  and sparse arrays. The corresponding range and angle recovery errors are shown in Fig.~\ref{fig:error}.

From Fig.~\ref{fig:rates}, we observe that for high SNRs, the proposed method with OMP-based recovery achieves the same hit rate as the classical processing for the full array with same false alarm rates. However, the full array consists of $12$ ($4$ Tx+ $8$ Rx) antenna elements, while the sparse array requires only half of these elements. On the other hand, reducing the transmitter and receiver elements drastically degrades the detection ability of classical-FFT processing. Interestingly, at lower SNRs, the hit rate of the classical processing (full array) reduces due to the side-lobe effect discussed earlier. Note that at high SNRs, the false peaks from the side-lobes are not prominent compared to the actual target peaks. On the contrary, the proposed method maintains the same hit rate with varying noise levels. The SOMP-based angle recovery in the proposed method further helps to increase the detection probability with a reduced false alarm rate, compared to OMP-based recovery. SOMP improves the detection ability by exploiting the correlation among the measurements across different pulses to recover the true target AOAs. In Fig.~\ref{fig:error}a, we observe that the classical method has a lower range recovery error for both full and sparse arrays, because of the refined range FFT computed in the classical method. The proposed method achieves a slightly higher range error of about $0.15$ m. Similarly, in Fig.~\ref{fig:error}b, the classical method slightly outperforms the proposed method in terms of angle recovery error. However, the classical method's angular resolution (hence, the error) depends on the array aperture. A higher angular resolution requires an increase in the array aperture and hence, the number of antenna elements. On the other hand, the proposed method's angular resolution is determined by the number of grid points $G$. Hence, the angle recovery error of the proposed method can be reduced with a finer grid $\bm{\phi}_{1\leq g\leq G}$. However, the number and locations of the antenna elements still affect the dictionary matrix $\mathbf{C}$, which in turn, determines the recovery probability of the CS-based algorithms.

\section{Summary}
\label{sec:summary}
We have proposed a novel sparse-recovery-based multi-target detection algorithm in the range-angle domain for MIMO FMCW radar. The proposed method enables a random array MIMO system to localize multiple targets in a sparse scene with reduced antenna elements compared to the traditional full array system. For range detection, we considered a DFT-based focusing operation with binary integration across pulses and virtual array channels. The binary integration in range detection provided a reduced missed detection probability than the classical non-coherent range-FFT processing. Finally, we considered a sparse recovery framework for target AOAs detection using both SMV and MMV-based CS recovery algorithms. Through numerical simulations, we illustrated the proposed method's target recovery compared to classical-FFT processing. Our numerical experiments suggest that the proposed method can achieve the traditional full-array hit rate with limited antenna elements. Furthermore, the MMV-based angle recovery can outperform both SMV-based and classical-FFT methods.

\bibliographystyle{IEEEtran}
\bibliography{references}

\end{document}